\documentstyle[prl,aps,epsfig,twocolumn]{revtex}

\begin{document}

\draft

\title{Structure and Magnetism of well-defined cobalt nanoparticles 
embedded in a niobium matrix}

 \author{M. Jamet$^1$, V. Dupuis$^1$, P. M\'elinon$^1$, G. Guiraud$^1$, A. P\'erez$^1$, W. Wernsdorfer$^2$, A. Traverse$^3$, B. Baguenard$^4$}
  
\address{$^1$ D\'epartement de Physique 
des Mat\'eriaux, Universit\'e Claude Bernard-Lyon 1 et CNRS, 69622 
Villeurbanne, FRANCE. \\
 $^2$ Laboratoire Louis N\'eel, CNRS, 38042 
Grenoble, FRANCE.  \\
$^3$ LURE, CNRS CEA MENJS, B\^at. 209A, BP 34, 91898 Orsay, FRANCE. \\
$^4$ Laboratoire de Spectrom\'etrie ionique et mol\'eculaire, Universit\'e Claude Bernard Lyon 1 et CNRS, 69622 Villeurbanne, FRANCE}

\maketitle

\begin{abstract}

Our recent studies on Co-clusters embedded in various matrices reveal that 
the co-deposition technique (simultaneous deposition of two beams : one for 
the pre-formed clusters and one for the matrix atoms) is a powerful tool 
to prepare magnetic nanostructures with any couple of materials even 
though they are miscible. We study, both sharply related, structure and 
magnetism of the Co/Nb system. Because such a heterogeneous system needs 
to be described at different scales, we used microscopic and macroscopic 
techniques but also local selective absorption ones. We conclude that our clusters are 3 nm diameter f.c.c truncated octahedrons with a pure 
cobalt core and a solid solution between Co and Nb located at the 
interface which could be responsible for the magnetically 
inactive monolayers we found.
The use of a very diluted Co/Nb film, further lithographed, would allow us 
to achieve a pattern of microsquid devices in view to study the magnetic 
dynamics of a single-Co cluster. 

\end{abstract}


\narrowtext

\section {Introduction}

Structural and magnetic properties of clusters, $\textit{i.e.}$ particles containing from
two to a few thousand atoms, are of great interest nowadays. From a 
technological point of view, those systems are part of the development of 
high density magnetic storage media, and, from a fundamental point of 
view, the physics of magnetic clusters still needs to be investigated. 
Indeed, to perform stable magnetic storage with small clusters, one has to 
control the magnetization reversal process (nucleation and dynamics), and 
thus make a close connection between structure and magnetic behavior. To 
reach the magnetic properties of small clusters, there are two 
available approaches : "macroscopic" measurements (using a Vibrating 
Sample Magnetometer (VSM) or a Super Quantum Interference Device (SQUID)) 
on a cluster collection (10$^{9}$ particles) that implicate statistical 
treatments of the data, and "microscopic" measurements on a single 
particle. From now, micro-magnetometers (MFM,~\cite{Chan93} Hall 
micro-probe,\cite{Geim97} or
classical micro-squid\cite{Wern97}) were not sensitive enough to perform magnetic 
measurements on a single cluster. The present paper constitutes the 
preliminary study toward magnetic measurements on a small single cluster 
using a new microsquid design. We focus on and try to connect structural 
and magnetic properties of a cluster collection. With a view to clear up 
structural questions, we first study the structure of nanocrystalline 
Co-particles embedded in a niobium matrix by means of Transmission 
Electron Microscope (TEM) observations, X-ray diffraction and absorption 
techniques. Then magnetization measurements are performed on the same 
particles to deduce their magnetic size and their anisotropy terms.

\section {experimental devices }

	We use the co-deposition technique recently developed in our laboratory 
to prepare the samples.\cite{Pare97} It consists in two independent beams reaching at 
the same time a silicon (100) substrate at room temperature : the 
pre-formed cluster beam and the atomic beam used for the matrix. The 
deposition is made in a Ultra High Vacuum (UHV) chamber 
(p=$5.10^{-10}$ Torr) 
to limit cluster and matrix oxidation.
	The cluster source used for this experiment is a classical laser 
vaporization source improved according to Milani-de Heer 
design.\cite{Mila90} It 
allows to work in the Low Energy Cluster Beam Deposition (LECBD) regime : clusters 
do not fragment arriving on the substrate or in the 
matrix.\cite{Pere97} The 
vaporization Ti:Sapphire laser used provides output energies up to 300 mJ 
at 790 nm, in a pulse duration of 3 $\mu$s and a 20 Hz repetition rate. It 
presents many advantages described elsewhere,\cite{Pell94} such as adjustable high 
cluster flux. The matrix is evaporated thanks to a UHV electron gun in 
communication with the deposition chamber. By monitoring and controlling 
both evaporation rates with quartz balances, we can continuously adjust 
the cluster concentration in the matrix.
We previously show that this technique allows to prepare nanogranular films 
from any couple of materials, even two miscible ones forbidden by the phase diagram 
at equilibrium.\cite{Negr99} We determine the crystalline structure and the morphology 
of cobalt clusters deposited onto copper grids and protected by a thin 
carbon layer (100 $\AA$). From earlier High Resolution Transmission Electron 
Microscopy (HRTEM) observations, we found that cobalt clusters form 
quasi-spherical nanocrystallites with a f.c.c structure, and a sharp size 
distribution.\cite{Tuai97,Pare97}
	In order to perform macroscopic measurements on a cluster collection 
using surface sensitive techniques, we need films having a 5-25 nm 
equivalent thickness of cobalt clusters embedded in 500 nm thick niobium 
films. We chose a low cluster concentration (1-5 $\%$) to make structural and 
magnetic measurements on non-interacting particles. One has to mention that 
such concentration is still far from the expected percolation threshold (about 
20 $\%$).\cite{Pare97}
	From both X-ray reflectometry and grazing X-ray scattering measurements, 
we measured the density of the Nb films: 92 $\%$ of the bulk one, and a 
b.c.c polycrystalline structure as reported for common bulk.
X-ray absorption spectroscopy (XAS) was performed on D42 at the LURE facility in 
Orsay using the X-ray beam delivered by the DCI storage 
ring\cite{Baud93} at the Co 
K-edge (7709 eV) by electron detection at low temperature (T=80 K)\cite{Mima94}. The porosity of the matrix is low 
enough and avoids the oxidation of the reactive Co clusters as shown in 
X-ray Absorption Near Edge Structure (XANES) spectra at the Co-K 
edge where no fingerprint of oxide on cobalt clusters embedded in 
niobium films is observed. The results of the Extended X-ray Absorption Fine Structure 
(EXAFS) simulations reveal the local distances between first Co-neighbors 
and their number for each component. 
Magnetization measurements on diluted samples were performed using a 
Vibrating Sample Magnetometer (VSM) at the Laboratoire Louis N\'eel in 
Grenoble. Other low temperature magnetization curves of the same samples 
were obtained from X-ray Magnetic Circular Dichroism (XMCD) signal. The 
measurement was conducted at the European Synchrotron Radiation Facility 
in Grenoble at the ID12B beamline. The degree of circular polarization was 
almost 80 $\%$, and the hysteresis measurements were performed using a 
helium-cooled UHV electromagnet that provided magnetic fields up to 3 
Tesla.

\section {structure}

The origin of the EXAFS signal is well established as mentioned in various 
references.\cite{Carr90} If multiple scattering effects are neglected on the first 
nearest neighbors, the EXAFS modulations are described in terms of 
interferences between the outgoing and the backscattered photoelectron wave 
functions. We use Mc-Kale tabulated phase and amplitude shifts for all types 
of considered Co-neighbors.\cite{Kale86} The EXAFS analysis is restricted to solely 
simple diffusion paths from the standard fitting code developed in 
the Michalowicz version\cite{Mich97} where an amplitude reduction factor 
S$_{0}^{2}$ equal to 0.7\cite{MRoy97} and an asymmetric distance distribution based on hard 
sphere model\cite{Prou97} are introduced. The first consideration traduces the possibility of 
multiple electron excitations contributing to the total absorption 
coefficient reduction. The 
second one is needed to take into account the difference between the core 
and the interface Co-atom distances in the cluster.\cite{Tuai97} So, in 
the fit, R$_{j}$  and s$_{j}$ values, corresponding to the shortest distance and to 
the asymmetry parameter of the j$^{th}$ atom from the excited one 
respectively, replace the average distance in 
the standard EXAFS formulation. We also define N$_{j}$ the coordination number, $\sigma_{j}$ the Debye-Waller factor  
of the j$^{th}$ atom, k the photoelectron momentum and $\Gamma$(k) its mean free path.
Structural parameters ($N_{j}, R_{j}, \sigma_{j}, s_{j}$) were determined from the 
simulation of the EXAFS oscillations (Fig. 1). As for some systems with 
two components (for example in metallic superlattices previously 
studied\cite{Baud93,Vdup93}), the first Fourier transform peak of the EXAFS 
spectrum presents a shoulder which can be understood unambiguously in terms of 
phase-shift between Co and Nb backscatterers for k values around 
5$\AA^{-1}$. 
This splitting in the real space corresponds to a broadening of the second 
oscillation in the momentum space\cite{Baud93} (see Fig. 1). Thus, in the simulations, we first consider two kinds of Co-neighbors : cobalt and niobium. But, a preliminary study of cobalt and niobium core levels by X-ray photoelectron spectroscopy reveals a weak concentration of oxygen inside the sample owing to the UHV environment. The core level yielding provides an oxygen concentration of about 5 $\%$. Such Co-O bonding is taken into account for the fit improvement. Moreover, from HRTEM and X-ray 
diffraction patterns \cite{Pare97,Tuai97}, we know :  
the mean size of the clusters (3 nm), their inner f.c.c 
structure with a lattice parameter close to the bulk one and their shape close to the Wulff
equilibrium one (truncated octahedron). Finally, the Co/Nb system can 
be usefully seen as a cobalt core with the bulk parameters and a more 
or less sharp Co/Nb interface. From these assumptions, we use the simulation of EXAFS oscillations to describe the Co/Nb interface and to verify it is relevant with an observed alloy in the phase diagram (tetragonal Co$_{6}$Nb$_{7}$).
The best fitted 
values of EXAFS oscillations are the following : \\ 
	- 70 $\%$ of Co atoms are surrounded with cobalt neighbors in the f.c.c phase 
with the bulk-like distance (d$_{Co-Co}$=2.50 $\AA$), corresponding 
to N$_{1}$=8.4, 
R$_{1}$=2.495 $\AA$, $\sigma_{1}$=0.1 $\AA$, s$_{1}$=0.18 $\AA$ in EXAFS 
simulations.   \\
	- 26 $\%$ of Co atoms are surrounded with niobium neighbors in the tetragonal 
Co$_{6}$Nb$_{7}$ phase, corresponding to N$_{2}$=3.1, R$_{2}$=2.58 $\AA$, 
$\sigma_{2}$=0.16 $\AA$, s$_{2}$=0.06 $\AA$ 
in EXAFS simulations.   \\
	- 4 $\%$ of Co atoms are surrounded with oxygen neighbors with a distance 
equal to d$_{Co-O}$=2.0 $\AA$ based on the typical oxygen atomic radii in chemisorption systems\cite{Drey99} or transition metal oxides. This environment 
corresponds to N$_{3}$=0.5, R$_{3}$=1.9 $\AA$, $\sigma_{3}$=0.04 
$\AA $, s$_{3}$=0.1 $\AA$ in EXAFS 
simulations.  \\
According to Ref.\cite{Hard69}, a 3 nm-diameter f.c.c truncated octahedron consists of 35.6 $\%$ core atoms (zone a), 27 $\%$ atoms in the first sublayer (zone b) and 37.6 $\%$ atoms in the surface layer (zone c). Let us propose the following compositions : a pure f.c.c Co phase in zone a, a Co$_{4}$Nb phase in zone b, and a Co$_{6}$Nb$_{7}$O$_{2}$ phase in zone c (i.e. : at the cluster-matrix interface). The corresponding coordination numbers : N$_{1}$(Co-Co)=8.5, N$_{2}$(Co-Nb)=3.1 and N$_{3}$(Co-O)=0.4 are in good quantitative agreement with the coordination numbers N$_{1}$, N$_{2}$ and N$_{3}$ we obtain from EXAFS simulations.
	Concerning the other fitting parameters, what is found is the high value for the mean free path of the photoelectron 
	($\Gamma =1.6$) 
and the Debye Waller factor for Co-metal environment ($\sigma$$>$ 
0.1 $\AA$). Notice 
that because we did not dispose of experimental phase and amplitude, but 
calculated ones, a large difference between sample and reference is 
expected, so their absolute values do not represent physical reality but 
only are necessary to attenuate the amplitude of oscillations. On the 
contrary, the total number of neighbors is fixed by TEM experiments which 
reveal a f.c.c-phase for the Co-clusters (so N$_{1}$+N$_{2}$+N$_{3}$=11$\pm$1). To follow the shape, position and relative amplitudes of the 
oscillations, N$_{j}$ is a free parameter for each 
component in the simulation and besides is related to the concentration 
of the j$^{th}$ atom from the Co-absorber one in the sample. This study 
finally evidences a diffuse interface between cobalt and 
niobium mostly located on the first monolayer.

In summary, we made a consistent treatment of all the experimental results obtained from different techniques. 
We notice that EXAFS spectra show unambiguously a smooth interface between 
miscible elements as cobalt 
and niobium. This information will be of importance and is the key to 
understand the magnetic behavior discussed below. 

\section{magnetism}

	Here, we present the magnetic properties of these nanometer sized 
clusters embedded in a metallic matrix. Furthermore such a system 
will be used to perform microsquid devices in order to reach magnetization 
measurements on an isolated single domain cluster. The present study deals with 
macroscopic measurements performed on a particle assembly (typically 
10$^{14}$) of cobalt clusters in a niobium matrix to describe the magnetic properties 
of the Co/Nb system. Because of the goal mentioned above, we focus on very 
diluted samples (less than 2 $\%$ volumic for Co concentrations). 
For these low cluster concentrations, magnetic couplings between particles are negligible whereas 
dipolar and RKKY interactions in the case of metallic matrix, 
are considered. Nevertheless, both last contributions which vary as 
1/d$_{ij}^{3}$ (where d$_{ij}$ is the mean distance between particles) 
are expected to be weak compared 
to ferromagnetic order inside the cluster. In a first approximation, 
we neglect any kind of surface disorder so that a single domain cluster can be seen as an 
isolated macrospin with uniform rotation of its magnetization. It means 
that the atomic spins in the cluster remain parallel during the cluster magnetization rotation. In an external applied field, the magnetic energy of a nanoparticle 
is the sum of a Zeeman interaction (between 
the cluster magnetization and the local field), and anisotropy terms (as shape, magnetocrystalline, surface 
(interface in our case) or strain anisotropy).
At high temperatures (T$>$100 K), anisotropy contributions of nanometric 
clusters can be neglected  compared to the thermal activation 
($K_{eff}V/k_{B}\approx 30 K$\cite{Comm99}) 
and clusters act as superparamagnetic independent entities. A way to 
estimate the interparticle interactions is to plot 1/$\chi$ vs. T in the 
superparamagnetic regime. 1/$\chi$ follows a 
Curie-Weiss-like law :
\begin{equation}
\frac{1}{\chi} =C(T-\theta)
\end{equation}
and $\theta$ gives an order of magnitude of the particle 
interactions.\cite{Dorm97} From 
experimental data, we give 1/$\chi$ vs. T on Fig. 2, and find $\theta$=1-2 K, which is 
negligible compared with the other energies of the clusters. 
In the superparamagnetic regime, we also estimate in Section A 
the magnetic size distribution of the clusters.
At low temperatures, clusters have a ferromagnetic behavior due to the 
anisotropy terms. And, in Section B, we 
experimentally estimate their mean anisotropy constant. 

\subsection{Magnetic size measurement}

 In the following, we make the approximation that the atomic magnetic moment is equal to 1.7 $\mu _{B}$ at any 
temperature (or 1430 emu/cm$^{3}$ like in the bulk h.c.p cobalt). 
Besides, our synthesized cobalt clusters have approximately a 3 nm diameter and contain at least 
1000 atoms. According to references \cite{Bill94,Resp98}, a magnetic moment 
enhancement only appears for particles containing less than 500 atoms. So 
in our size range we can assume that the atomic cobalt moment is close to 
the bulk phase one (m$_{Co}$=1.7$\mu _{B}$). 
We consider a log-normal size distribution :
\begin{equation}
f(D)=\frac{1}{D\sqrt{2\pi\sigma^{2}}}exp\Biggl(-\biggl(ln\Bigl(\frac{D}{D_{m}}\Bigr)\biggr)^{2}
\frac{1}{2\sigma^{2}}\Biggr) 
\end{equation}
where D$_{m}$ is the mean cluster diameter and $\sigma$ the dispersion. In the superparamagnetic regime,
 we can use a classical Langevin function $L(x)$ and write :
\begin{equation}
\frac{m(H,T)}{m_{sat}}=\frac{\int _{0}^{\infty}D^{3}L(x)f(D)\,dD}{\int 
_{0}^{\infty}D^{3}f(D)\,dD}, x=\frac{\mu_{0}H(\pi D^{3}/6)M_{S}}{k_{B}T}
\end{equation}
where H is the applied field ($\mu_{0}$H in Tesla), T the temperature and 
m$_{sat}$ 
the saturation magnetic moment of the sample estimated on magnetization 
curves at low temperatures under a 2 Tesla field. First of all, on Fig. 3, 
one can see that for T$>$100 K, m(H/T) curves superimpose according to Eq. 
(3) for a magnetic field being applied in the sample plane (we checked 
that the results are the same for a perpendicular applied field). 
Secondly, one can notice that for T=30 K, the magnetization deviation to 
the high temperature curves comes from the fact that the anisotropy is not 
negligible anymore, and one has to use a modified Langevin function in the 
simulation.\cite{Mull73} In this equation, we also assume the particles to feel the 
applied field, actually, they feel the local field which is the sum of the 
external field and the mean field created by the surrounding particles in 
the sample.
Furthermore, in the superparamagnetic regime, we fit experimental m(H,T) 
curves obtained from VSM measurements to find D$_{m}$ and $\sigma$, the mean diameter 
and dispersion of the "magnetic size" distribution, respectively (see Fig. 
4). For those fits, we still use the M$_{S}$ bulk value (the use of other ones 
given in references \cite{Brun89,Alde92,Dora97} leads quite to the same results (with an 
error less than 5 $\%$), the determining factors being D$_{m}$ and 
$\sigma$). Figure (5) 
displays D$_{m}$ and $\sigma$ for two niobium deposition rates (V$_{Nb}$=3 
$\AA$/s and V$_{Nb}$=5 $\AA$/s, 
respectively). Such results are compared with the real cluster sizes 
deduced from TEM observations. The magnetic domain is always smaller than 
the real diameter. Furthermore, the magnetic domain decreases as the 
deposition rate increases. This indicates that the kinetics of the 
deposition plays a crucial role for the nature of the interface. For 
example, we found a magnetic domain size of 2.3 nm (resp. 1.8 nm) for a 
3 nm diameter cluster when V$_{Nb}$=3 $\AA$/s (resp. 
V$_{Nb}$=5 $\AA$/s), the dispersion $\sigma$=0.24 remained the same.

\subsection{Anisotropy}

  The bulk value of the f.c.c cobalt cubic magnetocrystalline 
  anisotropy constant is : K$_{MA}$=2.7.10$^{6}$ 
erg/cm$^{3}$\cite{Chen94} less than the h.c.p bulk phase one
(4.4.10$^{6}$ erg/cm$^{3}$). The shape anisotropy constant K$_{shape}$ can be 
calculated from 
the demagnetizing factors and the saturation magnetization. 
 In case of weak distortions in the sphericity, the shape anisotropy for a 
prolate spheroid can be expressed as follows : 
\begin{equation}
E_{shape}=\frac{1}{2}\mu_{0}M_{S}^{2}(N_{z}-N_{x})\cos^{2}(\theta)=K_{shape}\cos^{2}(\theta)
\end{equation}
M$_{S}$ is the saturation magnetization of the particle : M$_{S}$ =1430 
emu/cm $^{3}$, $\theta$ 
the angle between the magnetization direction and the easy axis, and 
N$_{x}$, 
N$_{z}$ the demagnetizing factors along x-axis and z-axis respectively.
 We plot on Fig. 6, the constant anisotropy K$_{shape}$ as a function of the prolate 
spheroid deformation c/a \cite{Ahar96} (with c and a representing the wide and  
small ellipsoid axis, respectively). For a truncated octahedron, the ratio 
c/a has been evaluated lower than 1.2 which restricts the K$_{shape}$ value of the 
order of 10$^{6}$ erg/cm$^{3}$. However, we have no information about 
the magnitude of interface and strain anisotropies in our system.   \\
 Let us now experimentally evaluate the anisotropy constant K$_{eff}$ of 
cobalt clusters from low temperature measurements. Hysteresis curves are obtained from VSM experiments, but at very low temperatures ($\textit{i.e.}$ T$<$8 K), superconducting fluctuations appear due to the niobium matrix and prevent any magnetization measurements on the whole sample. So, we also use X-ray 
Magnetic Circular Dichroism as a local magnetometer by recording 
the MCD signal at the cobalt L$_{3}$ white line as a function of the applied 
magnetic field (for details on the method see Ref.\cite{Chen93}). The angle of the 
incident beam is fixed at 55$^{\circ}$ with respect to the surface normal and the magnetic field is parallel to the sample surface. The 
absorption signal is recorded by monitoring the soft X-ray fluorescence 
yield chosen for its large probing depth (1000 $\AA$). Finally, from hysteresis curves given by both VSM and XMCD techniques, we deduce m$_{r}$(T), the remanent magnetic moment vs. T down to 5.3 K, and we normalize it by 
taking : m$_{r}$(8.1K)$_{VSM}$=m$_{r}$(8.1K)$_{XMCD}$, the curve 
m$_{r}$(T)/m$_{r}$(5.3K) is given on Fig. 
7. 
To evaluate m$_{r}$(T), one can write :
\begin{equation}
m_{r}(T)=\frac{m_{sat}}{Cte}\frac{\int _{D_{B}(T)}^{\infty} 
D^{3}f(D)\,dD}{\int _{0}^{\infty} D^{3}f(D)\,dD}
\end{equation}
where D$_{B}$(T) is the particle blocking diameter at temperature T. Cte is a 
parameter independent of the particle size. Cte=$2$ if clusters have a 
uniaxial magnetic behavior and $3-\sqrt{3}$ if they have a cubic magnetic 
one. In order to rule out this Cte, we plot the ratio :
\begin{equation}
\frac{m_{r}(T)}{m_{r}(5.3K)}=\frac{\int _{D_{B}(T)}^{\infty} 
D^{3}f(D)\,dD}{\int _{D_{B}(5.3K)}^{\infty} D^{3}f(D)\,dD}
\end{equation}
One finds D$_{B}$(T) when the relaxation time of the particle is equal to the 
measuring time : $\tau=\tau_{0}exp(K_{eff}V/k_{B}T)=\tau_{mes}$.
\begin{equation}
D_{B}^{3}(T)=aT, a=\frac{6k_{B}}{\pi 
K_{eff}}ln\Bigl(\frac{\tau_{mes}}{\tau_{0}}\Bigr)
\end{equation} 
$\tau_{0}$ is the microscopic relaxation time of the particle, taken independent 
of the temperature. The fit result is presented on Fig. 7. We find 
a=3.5$\pm$0.1 
nm$^{3}$/K, and by taking $\tau_{mes}$=10 s, and 
$\tau_{0}$=10$^{-12}$-10$^{-9}$ s, we obtain K$_{eff}$=2.0 
$\pm$0.3.10$^{6}$ erg/cm$^{3}$.
By fitting Zero Field Cooled (ZFC) curves for different applied fields, we 
can also evaluate K$_{eff}$. Besides, if we neglect the blocked particle 
susceptibility, we have :
\begin{equation}
\frac{m_{ZFC}(H,T)}{m_{sat}}=\frac{\int_{0}^{D_{B}(H,T)}D^{3}L(x)f(D)\,dD}{\int_{0}^{\infty}D^{3}f(D)\,dD}
\end{equation}
Moreover, for low field values compared with the anisotropy field of 
cobalt clusters (estimated to be $\mu_{0}$H$_{a}$=0.4 T), we can make 
the approximation :
\begin{equation}
D_{B}^{3}(H,T)=af\Bigl(\frac{H}{H_{a}}\Bigr)T\approx a\Bigl(1+\alpha \frac{H}{H_{a}}\Bigr)T
\end{equation}
where a is the coefficient of Eq. (7), $\alpha$ a numerical constant. The ZFC 
curve fits are presented on Fig. 8. A linear extrapolation to 
$\mu_{0}$H=0 T 
also gives a$\approx$3.5 nm$^{3}$/K and an anisotropy constant of 
2.0$\pm$0.3.10$^{6}$ erg/cm$^{3}$ 
for the same numerical values as above. We found a similar result for the 
second sample with a niobium evaporation rate of 5 $\AA$/s. Finally, 
we experimentally found an anisotropy constant close to the one of 
quasi-spherical f.c.c cobalt clusters.

\section{discussion}

The "magnetic size" distribution is compared to the one obtained from 
TEM observations of pure Co-clusters prepared in the same experimental 
conditions (see Fig. 5(a)). For all the studied Co/Nb samples, we 
systematically find a global size reduction which might be related to the 
formation of a non-magnetic alloy at the interface as suggested by EXAFS 
simulations. The most significant parameter in the magnetically dead alloy 
thickness, seems to be the rate of deposition of the niobium matrix 
(V$_{Nb}$). 
As an example, we mention that for V$_{Nb}$=5 $\AA$/s, the reduction is twice the 
one for V$_{Nb}$=3 $\AA$/s (see Fig. 5(b)). That result suggests the model proposed 
in Fig. 9(a), 9(b).
As cobalt-niobium forms a miscible system\cite{Mass73}, we show that the more V$_{Nb}$ 
increases, the more the quantity of Nb-atoms introduced at the cobalt 
cluster surface increases. \\
To study the magnetism of the perturbed monolayers at the interface, we 
prepared a cobalt-niobium alloy using induction-heating under argon 
atmosphere with 40 $\%$-Co and 60 $\%$-Nb atomic weights. From classical X-ray 
$\theta$/2$\theta$ diffraction ($\lambda$=1.5406 $\AA$), we identified 
the $\beta$-phase given by the 
binary phase diagram : Co$_{6}$Nb$_{7}$. From VSM measurements on this sample, we 
found a remaining paramagnetic susceptibility $\chi$=10$^{-4}$ (for 2$<$T$<$300 K) 
corresponding to the "Pauli" paramagnetism of the sample. This feature 
could explain the "dead" layer at the cluster surface.
Obi and al.\cite{Yobi99} obtained two "magnetically dead" cobalt monolayers on 
cobalt-niobium multilayers evaporated by a rf-dual type sputtering method 
("magnetically dead" layers were also suggested by M\"uhge and 
al.\cite{Muhg97} for Fe/Nb multilayers). 
Finally, we can underline the fact that the pre-formed cobalt clusters by 
LECBD technique are very compact nanocrystallites which conserve a 
magnetic core even if embedded in a miscible matrix. 
The existence of a "magnetically dead" layer at the cluster-matrix 
interface may reduce surface effects compared with recent results obtained 
on smaller cobalt particles (150-300 atoms) stabilized in polymers.\cite{Resp98} The estimated mean
anisotropy constant might correspond to cubic magnetocrystalline 
or shape effects. To confirm this assumption, works are in progress to investigate the magnetic properties of a single cluster 
in a niobium matrix using a new microsquid technique. \\
One can also mention that for XMCD signals detected from the total 
electron yield method, the extraction of quantitative local magnetic 
values from the applicability of the individual orbital and spin sum rules 
is in progress.\cite{Chen95} Nevertheless, one can mention a small enhancement of the 
orbital/spin magnetic moment ratio.\cite{Chen95} Such increase might come from the 
orbital magnetic moment enhancement expected for small particles. 
\cite{Dora97} Systematic XMCD studies on clusters assembled Co/X films 
should be performed on Si-protected layers under synchrotron radiation to 
confirm these results.

\section{conclusion}

We have shown that the magnetic properties of nanoparticles can be 
evaluated unambiguously if we know the size, the shape and the nature of 
the interface. This latter is given by EXAFS spectroscopy. We summarize 
the main results :  \\
	- the mean like-bulk Co-Co distance (d$_{Co-Co}$=2.50 $\AA$) concerns the 3/4 of 
the atoms (namely : the core atoms) \\
	- Co-Nb bonds are located on roughly one monolayer at the surface of the 
Co-clusters embedded in the Nb-matrix.  \\
 Even though this interface is 
rather sharp, it is of importance since the interface thickness is on the 
same order of magnitude than the cluster radius.
In addition, some magnetic properties were approached by different 
complementary techniques as VSM magnetometry (at temperatures higher than 
8 K) and XMCD signal detected by the fluorescence yield method (at 
temperatures from 5.3 K to 30 K) under a magnetic field. We show the good 
result coherence on the superimposed range (8 K$<$T$<$30 K) for both techniques 
probing the whole thickness of the sample. The main result is the 
possibility of a "magnetically dead" layer at the interface Co/Nb, to 
relate to the alloyed interface (from EXAFS measurements) and to the 
moderate anisotropy value (found around 2.10$^{6}$ erg/cm$^{3}$). To confirm this assumption and to 
understand the role of the interface on the anisotropy terms involved in 
so low dimension magnetic nanostructures, XMCD measurements at the 
Co-L$_{2,3}$ 
edge have to be provided on a Co/Nb bilayer stacking (alternating 2 
monolayers of Co and 2 monolayers of Nb) with the same Nb-deposition rates 
as in our systems.

\section{Aknowledgements}

The authors would like to thank M. NEGRIER and J. TUAILLON for fruitfull discussions, C. BINNS from the University of Leceister, 
United Kingdom and J.VOGEL from the Laboratoire Louis N\'eel at Grenoble, 
France for their help during the first XMCD tests on the ID12B line of N. 
BROOKES at the ESRF in Grenoble.


\begin{figure}

\caption{EXAFS spectrum obtained on the sample containing 5 $\%$ of cobalt 
clusters embedded in niobium ($\circ$ : experimental data, continuous line : 
simulation). }

\label{fig 1}
\end{figure}

\begin{figure}

\caption{Inverse of the sample susceptibility plotted versus the 
temperature T, making a linear extrapolation of this curve for high 
temperatures, one obtains, when 1/$\chi \rightarrow$ 0, an idea of the interaction 
temperature : $\theta$=1-2 K.}

\label{fig 2}
\end{figure}

\begin{figure}

\caption{In the superparamagnetic regime, m(H/T) curves superimpose for 
T$=$100, 200 and 300 K. ($+$ : 300 K, $\times$ : 200 K, $\diamond$ : 100 K). At 
T$=$30 K ($\bullet$ : 30 K), 
the anisotropy energy is no more negligible, and one has to use a modified 
Langevin function to fit the curve.}

\label{fig 3}
\end{figure}

\begin{figure}

\caption{ We use a classical Langevin function to fit experimental 
magnetization curves m(H) in the superparamagnetic regime ($\bullet$ : 
experimental data, continuous lines : fits). This allows us to deduce the 
mean particle diameter D$_{m}$ and the dispersion $\sigma$ of the 
"magnetic" size 
distribution considering a log-normal distribution for cobalt clusters.}

\label{fig 4}
\end{figure}

\begin{figure}

\caption{(a) TEM size distribution : D$_{m}$=3.0$\pm$0.1 nm, 
$\sigma$=0.24$\pm$0.01 and 
"magnetic" size distribution : D$_{m}$=2.3$\pm$0.1 nm, 
$\sigma$=0.24$\pm$0.01 for a niobium 
deposition rate V$_{Nb}$=3 $\AA$/s and a log-normal distribution, (b) TEM size 
distribution : D$_{m}$=3.2$\pm$0.1 nm, $\sigma$=0.25$\pm$0.01 and "magnetic" size 
distribution : D$_{m}$=1.8$\pm$0.1 nm, $\sigma$=0.25$\pm$0.01 for a niobium deposition rate 
V$_{Nb}$=5 $\AA$/s.}

\label{fig 5}
\end{figure}

\begin{figure}

\caption{The volumic shape anisotropy energy (K$_{shape}$ in erg/cm$^{3}$) is plotted 
(continuous line) vs. the particle deformation c/a assuming it is a 
prolate spheroid. We also report volumic magnetocristalline anisotropy 
energy for f.c.c and h.c.p cobalt.}

\label{fig 6}
\end{figure}

\begin{figure}

\caption{ Remanent magnetic moment plotted vs. the temperature T. The signal 
is first normalized writting : m$_{r}$(8.1K)$_{VSM}$=m$_{r}$(8.1K)$_{XMCD}$, and then we take 
m$_{r}$(5.3K)=1. We see that the continuous line curve fits both VSM ($\bullet$) and XMCD ($\circ$) measurements. From this fit, we can deduce the anisotropy constant K$_{eff}$.}

\label{fig 7}
\end{figure}

\begin{figure}

\caption{(a) ZFC curves (black dots) taken for 6 different applied fields : (a) 
0.002 T, (b) 0.005 T, (c) 0.0075 T, (d) 0.01 T, (e) 0.015 T, (f) 0.02 T. 
(b) Fits (continuous lines) allow to deduce af(H/H$_{a}$) 
(D$_{B}^{3}$(H,T)=af(H/H$_{a}$)T) and then 
the anisotropy constant K$_{eff}$.}

\label{fig 8}
\end{figure}

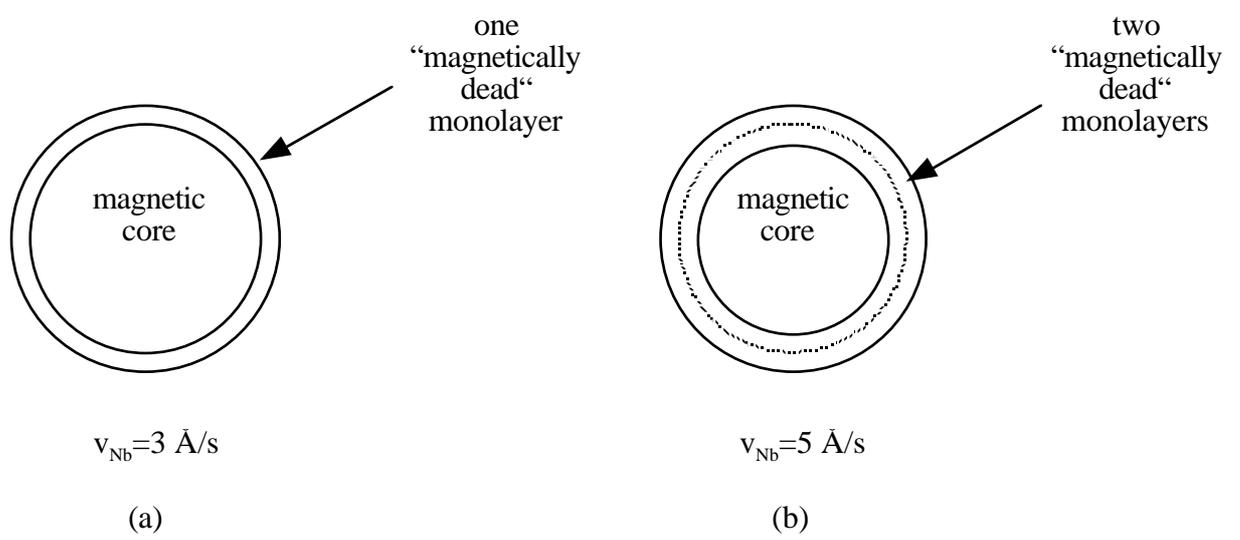
\begin{figure}

\caption{(a) Expected magnetic structure of cobalt clusters embedded in a 
niobium matrix, we find one 
"magnetically dead" cobalt monolayer (2$\times$3.5 $\AA$ in diameter) for a V$_{Nb}$=3 $\AA$/s deposition rate, (b) two "magnetically dead" cobalt 
monolayers (4$\times$3.5 $\AA$ in diameter) are found for V$_{Nb}$=5 $\AA$/s.}

\label{fig 9}
\end{figure}


\begin{thebibliography}{32}
\bibitem{Comm99}
The anisotropy constant used for this calculation is given in Section B.
\end{thebibliography}
\end{document}